\documentclass{ws-p8-50x6-00}
\def\rightnote{{ 
THE $\mu$-PARAMETER OF SUPERSYMMETRY 
}}
\def\mxth{\mathsurround=0pt }
\def\xversim#1#2{\lower2.pt\vbox{\baselineskip0pt \lineskip-.5pt
  \ialign{$\mxth#1\hfil##\hfil$\crcr#2\crcr\sim\crcr}}}             
\def\gtrsim{\mathrel{\mathpalette\xversim >}}                                    \def\lesssim{\mathrel{\mathpalette\xversim <}}  

\def\PLBold#1#2#3{{\it Phys.\ Lett.} {\bf#1B},  #3 (19#2)}
\def\IJMPA#1#2#3{{\it Int.\ J.\ of Mod.\ Phys.} {\bf A#1}, #3 (19#2)}
\def\MPLA#1#2#3{{\it Mod.\ Phys.\ Lett.} {\bf A#1}, #3 (19#2)}

\newcommand{\newc}{\newcommand}
\newc{\beq}{\begin{equation}}
\newc{\eeq}{\end{equation}}
\newc{\mpl}{\Lambda_{\rm P}}
\newc{\mgrav}{m_{3/2}}
\newc{\mgut}{\Lambda_{\rm U}}
\newc{\mstring}{M_{\rm string}}
\newc{\mw}{\Lambda_{\rm W}}
\newc{\msusy}{\Lambda_{\rm SUSY}}
\newc{\mint}{\Lambda_{\rm SUSY}}
\newc{\mmess}{\Lambda_{\rm M}}
\newc{\gev}{\,\mbox{GeV}}
\newc{\tr}{\mbox{Tr}\,}
\renewcommand{\bar}{\overline}
\newc{\Si}{\Sigma}
\newc{\eps}{\epsilon}
\newc{\ie}{{\it i.e.\/}}
\newc{\eg}{{\it e.g.\/}}
\newc{\gsim}{\lower.7ex\hbox{$\;\stackrel{\textstyle>}{\sim}\;$}}
\newc{\lsim}{\lower.7ex\hbox{$\;\stackrel{\textstyle<}{\sim}\;$}}

\begin{document}

\noindent

\phantom{a}     \hfill         
MIT--CTP--2923 \\
\phantom{a}     \hfill         
hep-ph/9911329          \\[3ex] 
\begin{center}

{\bf THE $\mu$-PARAMETER OF SUPERSYMMETRY}

\vspace{0.5cm}

{NIR POLONSKY}\\[1ex]

\vspace{0.3cm}

{\it
Center for Theoretical Physics, 
Massachusetts Institute of Technology, \\
77 Massachusetts Avenue, Cambridge, MA 02139-4307 USA 
\\E-mail: nirp@ctpgreen.mit.edu}

\end{center}
\vspace{1cm}

{\begin{center} ABSTRACT \end{center}}
\vspace*{1mm}
{\noindent
The Higgsino mass, or equivalently the $\mu$-parameter, 
plays an essential role in  determining the phenomenology of any 
supersymmetric model. Particularly, the size of the supersymmetry conserving
$\mu$-parameter must be correlated with the size of the soft supersymmetry
breaking parameters.
The source of this correlation in the underlying ultra-violet theory
is one of the mysteries of supersymmetry model building. 
The puzzle and the various possibilities 
for its resolution are reviewed, stressing
both phenomenological and theoretical aspects.
New proposals in the context of supergravity and gauge-mediation
frameworks for the soft supersymmetry breaking parameters are examined
in some detail.
}

\vspace*{2truecm}
{\begin{center} 
{Talk presented at SSS-99 \\
Supersymmetry, Supergravity, and Superstring \\
Seoul National University, Korea, June 23 -- 27, 1999}
\end{center}}

\vfill


\section{Introduction}\label{sec:s1}

The minimal supersymmetric extension (MSSM)
of the Standard Model (SM) of electroweak and strong interactions,
defined here by its minimal matter content, 
is described by the superpotential
\begin{equation}
W = -\mu H_{U}H_{D} - h_{U}H_{U}QU + h_{D}H_{D}QD+ h_{E}H_{D}LE,
\label{WMSSM}
\end{equation}
where $H_{U}$ (with hypercharge $Y = +1$) and $H_{D}$ ($Y = -1$)
are the two Higgs doublets required by the holomorphicity 
of $W$ and by consistency. Also, we employ the usual notation for 
the quark doublet ($Q$) and singlets ($U$ and $D$) and for the 
lepton doublet ($L$) and singlet ($E$). (Color, isospin 
and generation indices are suppressed.
For a tabulation of the different sign conventions and for a general review, 
see Ref.~1.) 
The field theory Lagrangain is given by the superspace integration
${\cal{L}} \sim \int d^{2}\theta W$.
Though the superpotential (\ref{WMSSM})
is that of the MSSM, it
provides the core superpotential terms of any extension
of the MSSM. (For a caveat, see Ref.~2.) 
In particular, the MSSM, or any of its extensions,
contains a (supersymmetric) Higgs mixing term $\mu H_{U}H_{D}$,
or more generally,
a Yukawa ``mass'' term involving a singlet or background 
superfield $X$, $W \sim \mu H_{U}H_{D} + \cdots \rightarrow 
W \sim \lambda X^{n} H_{U}H_{D} + \cdots$ and
 $\mu \rightarrow \lambda \langle X \rangle^{n}$
($n$ is determined by the dimensionality of the coupling $\lambda$,
$n = 0$ for $\lambda  = \mu$;
$n = 1$ if $\lambda$ is a yukawa term; and $n > 1$
in the case of non-renormalizable couplings.).

While the singlet interpretation will prove to be a convenient tool in the
discussion below, let us first assume 
a dimensionful parameter $\mu$ ($n = 0$)
in the effective theory which describes the regime between the
weak scale $\Lambda_{\rm W}$ and a few hundred GeV.
This assumption, which we
are about to justify, already hints at the puzzle we are would like to 
address: The natural choice of a scale for a dimesionful superpotential 
parameter is the scale of the ultra-violet theory, let it be the (reduced) 
Planck mass $\mpl \sim 2 \times 10^{18}$ GeV, the unification scale
$\mgut \sim 10^{16}$ GeV, or a messenger scale in a 
gauge-mediation framework $\mmess \gtrsim 10^{5}$ GeV.
Therefore, unless $\mu = 0$ (we return to and exclude this possibility below)
one expects $|\mu| \gg \Lambda_{\rm W}$. Nevertheless, a viable 
phenomenology requires $|\mu| \simeq \Lambda_{\rm W}$, or more correctly,
$|\mu| \simeq \Lambda_{\rm SSB}$,
where $\Lambda_{\rm SSB} \simeq (4\pi/h_{t})\Lambda_{\rm W}$
is the scale of the soft supersymmetry breaking (SSB) parameters
which is constrained from above by the stability of the weak scale
(\ie, the hierarchy problem), and $h_{t} \equiv h_{U_{3}}$ is the 
$t$-quark Yukawa coupling. 
This implies that all of the dimensionful
parameters of the supersymmetric extension have a similar origin,
even though from the low-energy point of view
they are of a very different nature. 
This puzzle was first formulated in Ref.~3, and
it suggests that the Higgs fields may be
distinguished from all other matter in the way in which they communicate
with heavy and supersymmetry breaking fields.

Supersymmetry provides technically natural
solutions to the hierarchy problem, \ie, 
the smallness of the weak scale $\Lambda_{\rm W} \simeq 0$ 
(in Planckian units). One can further understand,
using field-theory tools, spontaneous supersymmetry breaking in a hidden
sector at a given scale $\msusy$ 
as well as the propagation of the information to the observable 
sector and the generation of the SSB parameter scale $\Lambda_{\rm SSB}$, 
which in turn sets $\Lambda_{\rm W}$.
However, for the relation $\Lambda_{\rm SSB} \simeq \Lambda_{\rm W}$ to hold,
we must also understand the correlation $\mu \simeq \Lambda_{\rm SSB}$,
as we will attempt to do below.
Furthermore, this tension implies that $\mu$ cannot
be given by the ultra-violet cut-off scale but instead,
it parameterizes physics at that scale which is responsible for its 
smallness and its association with the SSB scale. This will lead us 
to treat $\mu$ as a spurion degree of freedom below, and adopt 
and generalize the ``singlet presentation'' given above,
but more importantly, it offers a potential benefit -- another 
door to the ultra-violet regime.

We must first argue, however, 
for the above conclusions in more detail.
This is done in Sec.~\ref{sec:pheno}.
In Sec.~\ref{sec:s2} we attempt to understand
the origin of the correlation between $\mu$ and $\Lambda_{\rm SSB}$.
We describe the various
possibilities in an effective operator language
and comment on their viability in the different frameworks.
The various manifestations of the $\mu$-problem are also discussed
and compared.
We then turn in Sec.~\ref{sec:s3} to describe two specific realizations
that were proposed recently in the context of supergravity and gauge-mediation
frameworks for the SSB parameters. In Sec.~\ref{sec:s4} we comment on
a generalization of the $\mu$-parameter to lepton number violating theories.
We conclude in Sec.~\ref{sec:s5}.

\section{Some Phenomenology}
\label{sec:pheno}

The (phenomenological) scale correlation between $\mu$ 
and the SSB parameters can be derived from
various considerations. For example, consider
the scalar potential given by
\begin{equation}
V = \sum\left|\frac{\partial W}{\partial \Phi}\right|_{\phi}
+ {\mbox{ gauge $D$-terms }} + V_{\rm SSB},
\label{V}
\end{equation}
where the summation
is over all chiral superfields $\Phi = H_{U},\,H_{D},\,Q,\cdots$,
$\Phi = \phi + \theta\psi +\theta^{2}F$ 
($\theta$ is the superspace coordinate), and the gauge $D$-terms
$\sim (g^{2}/2)\sum_{a,\,i}|\phi_{i}T^{a}\phi_{i}^{\dagger}|^{2}$, as usual.
The SSB terms contained in $V_{\rm SSB}$ are those terms that are mediated
in the SM visible sector by some messenger interactions which
communicate between the visible and some other hidden sector
in which supersymmetry is broken spontaneously. (It is said to be hidden
because by construction its interactions with the visible sector are strongly restricted, and in particular, could not be 
tree-level renormalizable interactions.)

The scalar potential (\ref{V}) 
includes the (tree-level improved) Higgs potential
\begin{eqnarray}
V(H) &=& (m_{H_{U}}^{2} + \mu^{2})|H_{U}|^{2} 
+ (m_{H_{D}}^{2} + \mu^{2})|H_{D}|^{2}  
- m_{H_{UD}}^{2}(H_{U}H_{D} + {\rm H.c.})  \nonumber \\
&+& \frac{g_{Y}^{2} + g^{2}}{8}
\left(|H_{U}|^{2} - |H_{D}|^{2}\right)^{2}
\label{VH}
\end{eqnarray}
where $g_{Y}$ ($g$) is the SM hypercharge ($SU(2)$) coupling.
Our notation does not distinguish a chiral superfield $\Phi$ and its 
scalar field component $\phi$, we assume real parameters, and 
$m_{H_{U}}^{2}$, $m_{H_{D}}^{2}$,  
and $m_{H_{UD}}^{2} \equiv B\mu$  are SSB parameters. 
Clearly, Max$[\mu^{2},\, m_{H_{U}}^{2},\, m_{H_{D}}^{2},\, m_{H_{UD}}^{2}]$
controls the scale of the potential and hence the realization and scale
of electroweak symmetry breaking (EWSB). The puzzle 
discussed above can now be rephrased: Why is the supersymmetry conserving
Higgs mixing parameter in the superpotential, $\mu$, 
of the same order of magnitude as the supersymmetry 
breaking mixing parameter in the scalar potential, $m_{H_{UD}}$?

Returning to the Higgs potential, it contains
a flat direction $m_{H_{U}}^{2} 
+ m_{H_{D}}^{2} + 2\mu^{2} - 2m_{H_{UD}}^{2} = 0$ 
(and hence, a light Higgs boson),
and is consistent with electroweak symmetry breaking iff
$(m_{H_{U}}^{2} + \mu^{2})(m_{H_{D}}^{2} + \mu^{2}) < |m_{H_{UD}}^{2}|^{2}$.
The latter condition is achieved in typical models 
radiatively as a result of  large Yukawa quantum corrections
(and hence, radiative symmetry breaking (RSB)). Once the symmetry is broken
then the weak scale, given by the $Z$ mass,  can be written in terms of
$\mu$ and the SSB parameters, or equivalently
\begin{equation}
\mu^2\ 
=\ 
\frac{m_{H_D}^2 - m_{H_U}^2 \tan^2\beta }
{\tan^2\beta -1} -  \frac{1}{2} m_Z^2, 
\label{MZ}
\end{equation}
where $\tan\beta =\langle H_U^0 \rangle /\langle H_D^0\rangle$ is the
usual ratio of Higgs vacuum expectation values (VEVs).
By observation, $|\mu|$ is given by a cancellation between the SSB parameters
and the experimentally determined weak scale, and hence, in order to avoid
fine tuning it must fall within this range. 
(The weak and SSB scales themselves are correlated by the 
cancellation of quadratic divergences and stability conditions.)
Diagonalization of the Higgs boson mass matrices, including EWSB effects,
in the $\mu \gg \Lambda_{\rm W}$ limit, exhibits a decoupling
of one Higgs (boson) doublet, which effectively does not participate
in EWSB, and of the Higgsinos, all with mass $\sim \mu$. 
The remaining light doublet boson which 
is responsible for EWSB
(and which contains the flat direction, and hence
a light Higgs boson) has no fermion superpartner. The resulting
quadratically divergent quantum correction again put an upper bound on 
the size of $\mu$. (In fact, one can use this observation
in order to calculate (leading) quantum correction
to the light-Higgs mass.\cite{decoupling})

It is worth commenting on the issue of fine tuning. No objective
definition for fine tuning exists
but clearly any sensible theory must avoid large 
cancellations -- unless they are a remnant of correlations or symmetries
in the ultra-violet theory (which may be mistaken by an infra-red
observer for fine tuning). 
Applying this mild criterion to electroweak
symmetry breaking Eq.~(\ref{MZ}) and recalling that the symmetry breaking
is encoded in the (one-loop renormalized) SSB parameters, imply that
the ultra-violet theory must correlate the SSB parameters and 
the effective low-energy $\mu$. 
When solving Eq.~(\ref{MZ}) in a given model,
one often finds a {\it de facto} correlation
between $\mu$ and the gluino mass $M_{\tilde{g}}$ 
which indirectly controls the renormalization of the SSB parameter
$m_{H_{U}}^{2}$.
Hence, in many cases correct phrasing of the fine-tuning issues need to be 
in terms of $|\mu/M_{\tilde{g}}|$, 
or more generally $|\mu/\Lambda_{\rm SSB}|$,
rather than in terms of $|\mu/m_{Z}|$ (as is often done). 
To reiterate, the issue is 
not the value of a precisely measured  infra-red parameter $m_{Z}$, 
but the understanding of the correlations
among the ultra-violet parameters which produce this value.
These correlations, of course, become more numerically constraining
and, therefore, more difficult to envision
or formulate as $\mu \sim \Lambda_{\rm SSB} \rightarrow \infty$ decouple.
(This limit corresponds to the restoration of the original hierarchy problem.)

Returning to the scalar potential (\ref{V}), 
it contains other terms involving $\mu$ which arise from
the cross terms in $|\partial W/\partial \Phi|^{2}$. These terms
constitute (non-holomorphic) tri-linear Higgs-left-right (LR) couplings
(where $\Phi_{L} = Q,\,L$ are the $SU(2)$ doublets and
$\Phi_{R} = U,\,D,\,E$ are the $SU(2)$ singlets).
After EWSB they provide chirality violating
off-diagonal LR entries in the sfermion $\tilde{f}$ mass-squared
matrices, 
\vspace*{0.3cm}
\begin{eqnarray}
\bar{m}^{2}_{\tilde{f}}
& = &
\left(\begin{tabular}{cc}
$m^{2}_{LL}$&$m^{2}_{LR}$\\
$m^{2}_{LR}$& $m_{RR}^{2}$
\end{tabular}\right),
\vspace*{0.3cm} 
\end{eqnarray}
where the LR mixing mass,
\begin{eqnarray}
m_{LR}^{2} &=& m_{f}\left( A_{f} - \mu\tan\beta\right)\,\,\,[{\rm or}\,\, 
m_{f}\left(A_{f} - \mu/\tan\beta \right) ],
\label{m2LR}
\end{eqnarray}
includes also SSB tri-liner $A$-parameters, which are implicitly
assumed to be proportional to the Yukawa coupling $\hat{A}_{f} =
y_{f}A_{f}$, which, in turn, is factored out,
and $\mu\tan\beta$ ($\mu/\tan\beta$)
terms appear in the down-squark and slepton mass matrices (up-squark
mass matrix). The requirement of a stable minimum and the positivity  
of the determinant constrain $|\mu|$ from above as well.
In addition, even if the EWSB minimum is stable, it may be only
a local minimum of the whole scalar potential (\ref{V}) while 
a charged and/or colored field acquires a non-vanishing VEV
along a direction in field space that corresponds to a
deeper (global) minimum, or along a flat direction. These considerations 
constrain the possible relations between the different parameters.
For example, for $h_{t} \sim 1$ the constraint 
\begin{equation}
(A_{t} \pm \mu)^{2} \leq 2(m^{2}_{Q_{3}} + m^{2}_{U_{3}})
\label{CCB}
\end{equation}
is found.\cite{CCB} 
(The undetermined sign on the left-hand side is given by sign$(A/\mu)$.)

All of the above establishes our previous assertion that a viable phenomenology
requires $\mu \sim \Lambda_{\rm SSB} \sim (1 - 10)\times \Lambda_{\rm W}$.
The crucial role of $\mu$ in determining the phenomenology
of the models, however, is apparent in many other cases. In particular,
the $\mu$ parameter also dominates (or contributes significantly to)
many ``supersymmetric quantum corrections'' 
either via the chirality flipping LR sfermion
mass squared or the Higgsino mass term. These include all the chirality
violating magnetic moment operators ($b\rightarrow s \gamma$ amplitude,
the anomalous muon magnetic moment $a_{\mu}$, and finite corrections
to fermion masses) as well as  radiative corrections to the light Higgs boson
mass (which are enhanced in the case of large LR stop mixing).
Most surprisingly, it also controls the one-loop threshold
corrections to gauge coupling unification predictions.\cite{unification}  
These, together with its role in determining the neutralino, chargino, 
and Higgs spectrum and couplings, imply that $\mu$ will be known
if supersymmetry is discovered and established experimentally. 

This leads to another question raised above: Is there sufficient evidence
(when assuming supersymmetry) that $\mu \neq 0$?
If it were zero it would resolve the most difficult part of the puzzle, 
\ie, why $\mu \simeq \Lambda_{\rm SSB}$, 
while  $\mu = 0$ can be understood in terms
of an enhanced (unbroken) symmetry (e.g. an $U(1)_{R}$ symmetry
under which the superpotential has non-trivial charge\cite{U1R}).
Setting a lower bound on a parameter rather than a mass of a physical state
is not trivial. However, it can be done in this case by consideration of chargino
pair production.
By observation, the (tree-level) chargino mass matrix,
\vspace*{0.3cm}
\begin{eqnarray}
\label{nchargino}
m_{\tilde{\chi}^{\pm}}
& = &
\left(\begin{tabular}{cc}
$M_{2}$ & $\sqrt{2}m_{W}\sin\beta$\\
$\sqrt{2}m_{W}\cos\beta$& $-\mu$ \\
\end{tabular}\right),  
\vspace*{0.4cm}
\end{eqnarray}
contains charginos degenerate in mass with the $W$ bosons for
$\mu = M_{2} = 0$ and $\tan\beta = 1$, where $M_{2}$ is a SSB 
Wino mass parameter.
Such charginos would have escaped 
detection at the $Z$-pole (LEPI), 
but would be pair-produced in abundance at the $WW$ threshold (LEPII).
They would then decay to a massless neutralino and a $W$ boson,
hence, imitating $WW$ production. However, no such deviations 
from SM predictions were observed, excluding this possibility.
In practice, one needs to consider neutralino production as well.
Surprisingly, a small region of parameter space with
$\mu \simeq M_{2} \simeq {\cal{O}}({\rm GeV})$ survived 
a careful analysis of all $Z$-pole data\cite{smallmu}
even when including radiative corrections and off-peak data.
While neutralino production above the $Z$-peak is complicated
by $t$-channel sneutrino exchange,\cite{ellisetal} the $WW$-threshold
constraints (even when applied to one on-shell and one slightly 
off-shell chargino) provide relatively a model-indepndent constraint
$|\mu| \gtrsim \Lambda_{\rm W}$, as suggested in Ref.~8.

Throughout the discussion it is assumed for simplicity
that $\mu$ is a real parameter and
we conveniently identify its ultra-violet and infra-red values.
In general $\mu$ is a complex parameter that carries a physical phase.
While it has been argued recently\cite{phases} 
that such a phase could be substantial
due to accidental cancellations among various contributions 
to, e.g.,  the neutron dipole moment, 
we will not entertain this possibility here.
Phases are also less constrained in the $U(1)_{R}$ limit\cite{phasesR} 
$|\mu| \ll m_{\tilde{f}}$, but again we do not consider this option here.
In practice, our discussion is mostly independent of any 
assumption about the phases. 
The sign of the real $\mu$ is still physical as it affects
the various quantum corrections and phenomena discussed above, but again
is of no concern in the discussion below. 
Sign($|\mu|$) is a renormalization group invariant. The magnitude $|\mu|$, 
on the other hand, is subject to wave function renormalization
so that the superpotential parameter $\mu$ is renormalized only in proportion
to itself, and only by electroweak and Yukawa couplings. 
In particular, it cannot mix via renormalization with the SSB parameters. 
Since our discussion  here is mostly qualitative we can 
safely ignore these effects which are typically small.

Equipped with (rough) upper and lower bounds on the absolute value
of the $\mu$-parameter, which suggest that it is correlated
with the SSB parameters, we turn to a discussion of the possible origin
of this correlation.

\section{Operator Analysis}
\label{sec:s2}

Hereafter we adopt the point of view that in the ultra-violet (UV) theory,
given its symmetries, $\mu_{UV} = 0$, while $\mu_{IR} \simeq \Lambda_{\rm SSB}$
is generated by terms which appear once supersymmetry is spontaneously
broken in some hidden sector of the theory.
Such terms break the relevant symmetries either explicitly or spontaneously 
in the effective infra-red (IR) theory. 
(As discussed above, 
$\mu_{IR}$ renormalization cannot change its order of magnitude
and is ignored.)
The parameter $\mu_{IR}$ then implicitly carries 
information on the symmetries
of the UV theory and on their violation by supersymmetry breaking.
It must also contain some information on the supersymmetry breaking fields.
This is seen most clearly once we return to viewing $\mu$ as a VEV of 
(a physical or an auxiliary component of) a (background) singlet superfield $X$
which may or may not participate directly in supersymmetry breaking.
More generally, let us denote by $X$ a singlet object that can couple
to the holomorphic bilinear $H_{1}H_{2}$, leaving aside for the moment its
identity or our definition of a singlet object. One can describe 
the superpotential and SSB
Higgs mixing parameters in terms of effective $F$-term ($\int d^{2}\theta$)
and $D$-term ($\int d^{2}\theta d^{2}\bar\theta$) operators
\begin{eqnarray}
{\rm SUSY \ Higgs \  mixing}\ (\mu): && 
a\int d^2 \theta\, H_1 H_2
\left[ \frac{X^{n}}{M^{n-1}} \right] 
\nonumber \\ 
{\rm and} &&
a^{\prime}\int d^2 \theta d^{2}\bar\theta \, H_1 H_2
\left[ 1 + \frac{X^\dagger}{M} +\cdots \right] + {\rm H.c.}, 
\label{mu}\\
{\rm SSB \ Higgs \ mixing} \ (m_{H_{UD}}^{2}): && 
b\int d^2 \theta\, H_1 H_2
\left[ X + \frac{X^{2}}{M} 
+ \cdots 
\right] 
\nonumber \\ 
{\rm and} &&
b^{\prime}\int d^2 \theta d^{2}\bar\theta \, H_1 H_2
\left[ \frac{X^{\dagger}X}{M^{2}} 
+ \cdots \right] + {\rm H.c.}, 
\label{m3}
\end
{eqnarray}
with the usual conventions $\int d^{2}\theta X = F_{X}$ in Eq.~(\ref{m3})
and $\int d^{2}\bar\theta X^{\dagger} = F_{X}^{\dagger}$
in Eqs.~(\ref{mu}) and (\ref{m3}).
The scale $M$ parameterizes the mediating interactions and 
is of the same order of magnitude as the scale of the mediation 
of supersymmetry breaking from the hidden to the visible sector $\msusy$.
It may be a true parameter or a background field itself 
$M = \langle Y \rangle$.
We wrote explicitly all the higher order 
operators in the $F$-term integral in Eq.~(\ref{mu}) since
only the scalar component
of $X$ contributes in this case, 
and for sufficiently large $X$ higher order operators
could be considered. 
Similarly, the leading $D$-term operator in (\ref{mu})
can also be identified with $\langle X \rangle/M \sim 1$.
Otherwise, higher order operators are denoted by $\cdots$.

To reach a resolution of the puzzle one 
needs to obtain the same order of magnitude
in both operators (\ref{mu}) and (\ref{m3}). (Note that in the latter case
there could be contributions from non-singlet fields $ZZ^{\dagger}$.) 
We now need to understand the nature of $X$, its value $\langle X \rangle$,
and its relation to supersymmetry breaking, \ie, $\langle F_{X} \rangle$.
We also need to understand the expectation for the coefficients, which
is crucial in distinguishing realizations in the different frameworks.
Finally, one may want to understand the symmetries which forbid/allow
the various operators. Note that the $\mu$ term breaks, in general,
both $U(1)$ Peccei-Quinn and $R$ symmetries while the SSB mixing term
breaks, in general, only the Peccei-Quinn symmetry. Both 
symmetries could play a role
in providing the appropriate selection rules for the operators.
In particular, $R$ symmetry is a powerful tool 
as it does not commute with supergravity ($R(\theta) = -1$) 
and all hidden and visible sectors are charged under it
and thus the resulting selection rules can be applied to also mixed 
hidden-visible operators.

Let us begin and examine the different operators. Assume that the operators
are tree-level operators and that the dimensionless
coefficients are generic
${\cal{O}}(1)$ coefficients. This corresponds to the 
``classical'' supergravity framework (\ie, tree-level gravity mediation 
of the SSB parameters in the visible sector via Planck suppressed operators), 
in which case $M \sim \mpl$
and $\Lambda_{\rm SSB}$ is simply given by the gravitino mass 
$m_{3/2} = \msusy^{2}/\sqrt{3}\mpl$ 
(assuming cancellation of the cosmological constant),
which, in turn, has to be fixed.
A primordial $\mu$ term ($n=0$) could be forbidden by $R$ symmetry\cite{R,NP1}
if $R(H_{U}H_{D}) \neq R(W) = 2$, for example.
It could also be forbidden by a variety of other symmetries such as
a Peccei-Quinn ($PQ$) symmetry\cite{NK,PQ} under which 
$PQ(H_{D}) +PQ(QD) = 0$ and $PQ(H_{U}) +PQ(QU)  = 0$,
a discrete $Z_{3}$ (or higher) symmetry\cite{nmssm},
a flavor\cite{flavor} $U(1)$, 
an extended gauge structure,\cite{prime}
or a grand-unified (GUT) symmetry.\cite{gut,gutmu}
(It could also be forbidden by modular invariance in string theory,
in which case the effective $\mu$ is a function of the 
moduli fields.\cite{string})
 
The next class of operators 
is the $n=1$ Yukawa operators in the superpotential
which is relevant for our purposes iff $\langle X \rangle \simeq m_{3/2}$.
The most advertised realization of this is the next to minimal MSSM (NMSSM)
in which the spectrum is extended by a gauge singlet and a discrete
$Z_{3}$ symmetry is imposed.\cite{nmssm} The singlet survives to low energies
and a mechanism similar to the one leading radiatively to EWSB with
$\langle H_{U,\,D} \rangle \sim \Lambda_{\rm SSB}$ now generates
$\langle X \rangle \sim \Lambda_{\rm SSB}$ (where the potential in stabilized
by a $X^{3}$ term in the NMSSM superpotential). 
While the supersymmetry conserving VEV of $X$ induces $\mu$,
the supersymmetry breaking VEV of the auxiliary component
$F_{X} \sim X^{2}$ provides the SSB mixing term 
(the first $F$-term operator in Eq.~\ref{m3}). 
It could also arise from higher order terms in the 
first ($F$-term) integral in Eq.~(\ref{m3}), which effectively
correspond to a SSB term $\sim AXH_{U}H_{D}$ in $V_{\rm SSB}$
if $X^{2} \rightarrow X_{0}X$ and $X_{0}$ is a supersymmetry breaking
(hidden-sector) field with $F_{X_{0}} \sim \msusy^{2} \sim m_{3/2}\mpl$.
While the extended spectrum is consistent
with all phenomenological constraints, the spontaneously 
broken $Z_{3}$ symmetry leads to post-inflationary domain-wall problem
which disfavors this construction.\cite{nmssm} 
A distinctive alternative is to gauge the extra $U(1)$ 
which appears\cite{prime}
(with $Q(X) = -Q(H_{U}H_{D})$), in which case the $Z_{3}$
is a harmless discrete subgroup. $X$ in this case is only a SM singlet
and not a gauge singlet.
Aside from the presence of
an additional neutral massive gauge boson, it is distinguished
from the usual NMSSM by the form of the quartic potential of the singlet
that now arises from the (extended) $D$-terms in (\ref{V}).
We will explore this option below in some detail, but in a different context.

A different approach is to consider a background field
$X$ that is decoupled from the sub-TeV theory but its 
VEV is sufficiently small so that terms in the low-energy 
theory could be proportional to it. For example, 
in GUTs the $\mu$-puzzle is only
an extension of the celebrated doublet-triplet problem,\cite{gut} 
\ie, the splitting of the {\bf{5}} and ${\bf \bar 5}$ of $SU(5)$ to a light
pair of $SU(2)$ doublets and a heavy pair of color $SU(3)$ triplets.
Assuming that such splitting occurs with exactly massless doublets
in the limit of global supersymmetry, 
once the ${\cal{O}}(m_{3/2})$ SSB 
terms for the heavy GUT fields are introduced, they explicitly break
the global supersymmetry and they can appropriately shift the doublet mass,
\ie, the $\mu$-parameter.\cite{gutmu}  
$X$ in this case is either a heavy singlet 
or a field in the adjoint representation of the GUT group.
A different and more recent proposal,\cite{paper1}
about which we will elaborate below, relates the ${\cal{O}}(m_{3/2})$
shift in $X$ and $F_{X}$ to a radiative generation of a tadpole term
for $X$ once supersymmetry is broken in the hidden sector.
$X$ in this case is a (total) singlet which decouples at a scale 
$\sim m_{3/2}\mpl$ but its VEVs are slightly shifted from zero
$X \sim m_{3/2} + \theta^{2}m_{3/2}^{2}$.

An alternative to renormalizable operators (and an option which is 
common in model building) is that the symmetries in the ultra-violet theory
allow only for non-renormalizable operator realization of $\mu$.
The leading $n= 2$ operators require  $X \simeq m_{3/2}\mpl$, 
the geometrical mean of the weak and Planck scale which also corresponds
(in the supergravity framework) to the scale of supersymmetry breaking 
in the hidden sector, $\msusy$. This scale also corresponds to the invisible
axion window and suggests that an anomalous symmetry such as a 
Peccei-Quinn\cite{NK,PQ} or a $R$-symmetry\cite{R,NP1} may dictate, in this 
case, the selection rules for the operators. $X$ is related to the breaking
of such a symmetry, though it could be in many cases a hidden sector field
(since its coupling is suppressed by powers of $\mpl$).
It is possible to relate the symmetry breaking
(particularly in the case of a $R$-symmetry)
to supersymmetry breaking in the hidden sector.
The SSB mixing parameters could arise from terms 
in the second ($D$-term) integral in Eq.~(\ref{m3}) 
with fields $X$ (and $Z$) corresponding 
to hidden sector fields with $F \sim m_{3/2}\mpl$.
The $n = 3$ case is also quite interesting, since one
could identify $X^{3} \equiv W$ 
where (assuming cancellation of the cosmological 
constant)  the superpotential VEV is given by that of the hidden 
supepotential\cite{CM} $\langle W_{hidden} \rangle  = m_{3/2}\mpl^{2}$,
which parameterizes in this case supersymmetry breaking.
Alternatively, $X^{3} \equiv W^{\alpha}W_{\alpha}$ could be identified
with a (hidden sector) gaugino condensate as long as
$\langle W^{\alpha}_{hidden}W_{\alpha\, hidden}\rangle =m_{3/2}\mpl^{2}$
as occurs in non-renormalizable hidden sector models.\cite{smallmu}
($W_{\alpha}$ is the chiral representation of the gauge superfield.)

The last possibility suggests that the hidden-visible gravity-mediated
mixing occurs in the holomorphic gauge kinetic function 
$f_{\alpha\beta} =
\delta_{\alpha\beta}(1 + (H_{U}H_{D}/M^{2}) + \cdots)/2g^{2}$ 
(which determines the gauge Lagrangian 
$f_{\alpha\beta}W_{\alpha}W_{\beta}$ 
and $g$ here is the relevant gauge coupling)
rather than in the superpotential.
(Note that the holomorphicity of the Higgs bilinear is explicitly exploited.)
Such a general form of the gauge kinetic function 
can also lead to a radiative but quadratically divergent 
(and thus, not negligible)
contribution to $\mu$ which is proportional to the SSB gaugino mass.\cite{QD}
The mixing could also occur in the non-holomorphic K\"ahler potential,
where again, one can exploit the holomorphicity of the Higgs bilinear 
and write\cite{GM} the terms which appear in the second ($D$-term)
integral in Eq.~(\ref{mu}). $X$ is a hidden sector field and
once it is integrated out 
$\int d^{2}\bar\theta X^{\dagger} \rightarrow F_{X}^{\dagger} \simeq
m_{3/2}\mpl$ then this term is reduced to a usual superpotential term.
(Note that similar $X^{\dagger}QQ^{\dagger}$
must be forbidden to avoid flavor non-diagonal sfermion masses,
as could be done by a $R$ symmetry.) In fact, in supergravity
(rather than the rigid supersymmetry of the effective theory which is 
implicitly assume in writing Eqs.~(\ref{mu}) and (\ref{m3})) 
it is sufficient to write a K\"ahler potential
which includes the holomorphic term
$K = H_{U}H_{D} + {\rm H.c.}$. Using a K\"ahler transformation
\vspace*{0.3cm}
\begin{center}
$K \rightarrow K -(H_{U}H_{D} + {\rm H.c.})$ 
and $W \rightarrow e^{(H_{U}H_{D}/\mpl^{2})}W$,
\end{center}
\vspace*{0.3cm}
and substituting $W = \langle W_{hidden}\rangle  + W_{\rm MSSM}$
where $\langle W_{hidden}\rangle \simeq m_{3/2}\mpl^{2}$
and $W_{\rm MSSM}$ is given by Eq.~(\ref{WMSSM}) but without the $\mu$-term,
reproduces the same result. Hence, one can identify this operator
with the first $n=3$ operator discussed above.
(Note, however, that the renormalizable and non-renormalizable
K\"ahler potential operators have a different holomorphy
structure in this case.)

The identity $m_{3/2} \simeq \Lambda_{\rm SSB}$ and the multitude of
available tree-level operators more than resolves the $\mu$
problem in supergravity and explains the correlation
$\mu \simeq \Lambda_{\rm SSB}$.
Some of the results are summarized in Table \ref{table:t1}.
\begin{table}[t]
\caption{The required field values for consistent generation of $\mu$
and possible symmetry sources for selection rules. In supergravity mediation
$\Lambda_{\rm SSB} \simeq m_{3/2}$.}
\begin{center}
\label{table:t1}
\begin{tabular}{|c|c|c|}
\hline
&&\\
$n$ & $\langle X \rangle$  & Symmetry\\ && \\ \hline
&& \\ 1 & $\Lambda_{\rm SSB}/a$ & $Z_{3}$, gauged $U(1)$, GUT\\
&& \\ 2 & $M\Lambda_{\rm SSB}/a$ & Peccei-Quinn, $R$\\
&& \\ 3 & $M^{2}\Lambda_{\rm SSB}/a$ & $R$ \\ && \\ \hline
\end{tabular}
\end{center}
\end{table}
Note that though one can clearly understand 
$\mu \sim \Lambda_{\rm SSB} \sim M_{\rm gluino}$, this relation
is not renormalization group invariant. The fine tuning issue
can then be phrased as understanding the special value $\mu/M_{\rm gluino}$ 
at the weak rather than the ultra-violet scale. The situation improves
in this respect in models with lower mediation scales, but as we shall see,
different complications arise in these cases.

It is possible that the sparticle spectrum contains
both multi-TeV sfermions (superpartners of the first and 
second family sfermions) and sub-TeV Higgs particles, gauginos, and
third family sfermions. The former ensure
decoupling of potential contributions
to sensitive flavor changing neutral currents observables, while
the latter allow for natural solutions to EWSB.\cite{heavies}
If the gravitino mass is still in the sub-TeV\cite{heavy1} 
then $\mu \sim m_{3/2}$
is sufficient. However, if the gravitino sets the scale for the heavy 
sector,\cite{heavy2} one must invoke a Peccei-Quinn 
(or other) symmetry (alongside the 
$R$-symmetry whose role is now to suppress the gaugino masses) 
in order to ensure sub-TeV Higgs mass parameters.

An alternative to gravity mediation is the gauge mediation framework
in which the mass-dimension one and two SSB parameters are induced
by gauge loops at one and two loop orders, respectively, so that
$m_{\rm gaugino} \sim m_{\rm sfermion} \sim \Lambda_{\rm SSB} \sim 
(\alpha/4\pi)\mmess \gg m_{3/2}$. Here, $\alpha$
is a properly renormalized SM gauge coupling 
and $\mmess \sim (4\pi/\alpha)\Lambda_{\rm SSB} 
\sim (4\pi/\alpha)(4\pi/h_{t})\Lambda_{\rm W} 
\sim 10^{4 - 6}$ GeV is a messenger scale, \ie, the mass scale of 
the messenger fields which
communicate between the SM gauge interactions and a low energy
supersymmetry breaking sector (the equivalent of the supergravity
hidden sector).   While supergravity interactions cannot be eliminated,
their effects are typically suppressed by the small gravitino mass $m_{3/2}
\sim \mmess^{2}/\mpl$.(For an exception, see Ref.~28.)
However, the mixing terms $\mu H_{U}H_{D}$ and $m_{H_{UD}}^{2} H_{U}H_{D}$
explicitly break the Peccei-Quinn symmetry and therefore
cannot arise from gauge loops. Instead, they could arise
from Yukawa loops if the messenger sector is extended appropriately.
(In fact, if $\mu \neq 0$ then 
$m_{H_{UD}}^{2} \sim \mu\Lambda_{\rm SSB}$ 
arises at two-loop order from one-loop (gauge) renormalization
once the gaugino mass is induced at one loop,\cite{noB} 
but this is immaterial
for our discussion here.) However, in the case of Yukawa interactions
one does not find a hierarchy similar to 
$m^{2}_{\rm sfermion}/m_{\rm gaugino}
\sim [(\alpha/4\pi)^{2}\mmess^{2}]/
[(\alpha/4\pi)\mmess] \sim \Lambda_{\rm SSB}$
which appears in the case of gauge loops.
Instead, one typically finds\cite{DGP}  
$m^{2}_{H_{UD}}/\mu \sim [(h/16\pi^{2})
\mmess^{2}]/[(h/16\pi^{2})\mmess] \sim \mmess$.
That is, both dimension one and two mixing parameters arise
at the same loop order.

The operator language is convenient for comparison with the supergravity case.
The $\mu$ parameter would arise from Eq.~({\ref{mu})
with $X \sim \sqrt{|F_{X}|}\sim \mmess$ given by the messenger scale and 
(assuming one-loop effects) $a \sim a^{\prime} \sim 1/16\pi^{2}$.
Similarly, the SSB $m_{H_{UD}}^{2}$ parameter
arises from Eq.~({\ref{m3}) with $b \sim b^{\prime} \sim 1/16\pi^{2}$.
(Some coefficients may vanish, depending on the exact structure one assumes.)
While in the case of tree-level supergravity operators
one has $m_{H_{UD}}^{2}\sim \mu\Lambda_{\rm SSB} \sim \mu^{2}$,
if both arise quantum mechanically at the same loop order then one has 
instead $m_{H_{UD}}^{2} \sim \mu \mmess$. 
As a result, a new hierarchy problem
that shadows the gauge mediation framework emerges.
One avenue to resolve the gauge-mediation variant of the $\mu$ puzzle
is to allow for $\mu$ and $m_{H_{UD}}^{2}$ generation at tree-level
and hence to reduce the problem to its ``supergravity form'', 
or alternatively
to allow for $m_{H_{UD}}^{2}$ generation only at higher loop level.
Both approaches require the introduction of dedicated singlets $X$ 
and of non-trivial structures and/or interactions.
A NMSSM realization of the former will be presented below.

It was recently proposed\cite{AM} 
that supergravity mediation may also arise
only at the quantum level with $\Lambda_{\rm SSB} \sim (\alpha/4\pi)m_{3/2}$
and the gravitino mass $m_{3/2} \gtrsim 10$ TeV, which is similar in size to 
the messenger scale of gauge mediation.
This is the anomaly mediation mechanism. 
There, the theory is assumed to preserve at all orders 
a geometrical separation of the hidden and observable sectors,
and supersymmetry breaking
effects in the observable  sector can arise only through the 
interactions of the supergravity multiplet. This leads to 
relations similar to those of gauge mediation
but with quite different and distinct coefficients. 
Specifically, the mediation of supersymmetry breaking
due to the supergravity multiplet can be extracted 
(in an appropriate gauge) by
introducing appropriate powers of a background field
$\phi = 1 + \theta^{2}m_{3/2}$, the holomorphic compensator,
to the different operators 
(so that the action is rendered Weyl invariant). 
The compensator, which in this parameterization is the only 
source of supersymmetry breaking in the observable sector, then allows for 
$\mu$ and $m_{H_{UD}}^{2}$ generation from the $D$-term integral
in Eq.~(\ref{mu}). However, the tree level operator
leads in this case to $\mu \sim m_{3/2} \gg \Lambda_{\rm W}$.
Even if one sets $a^{\prime} \ll 1$, 
the hierarchy problem described in the case of gauge mediation appears.
Thus, one must forbid tree-level generation and rely on radiative generation
(except, in principle,  in the case of the NMSSM with 
$\langle X \rangle \sim \Lambda_{\rm SSB}$).
Proposed solutions\cite{AMmu} have to rely on extended structures
and, furthermore, disturb the minimality which is the essence of
the initial proposals of this framework.

Lastly, we would like to point out a higher order $D$-term operator,\cite{bp}
\begin{eqnarray}
{\rm SSB \ Higgsino \  mixing}\ (\tilde\mu): && 
c\int d^2 \theta d^{2}\bar\theta \, DH_1 DH_2
\left[ \frac{XX^\dagger}{M^{3}} \right] + {\rm H.c.}, 
\label{softmu}
\end{eqnarray}
where $D = \partial/\partial\theta$ is the covariant superspace derivative
with mass dimension $[D] = 1/2$.
Its operation selects the Higgsino components of the Higgs superfields 
$H_{D,\,U}$ while the superspace integration leads to $\tilde\mu \sim
|F_{X}|^{2}/M^{3}$. Such a SSB Higgsino mass leads to the issue
of ``non-standard'' supersymmetry breaking terms\cite{NSSSB} 
as it can always be rotated into a combination of the usual $\mu$-term 
and non-holomorphic $A^{\prime}H^{\dagger}\phi_{L}\phi_{R}$
supersymmetry breaking terms ($\phi_{L,\,R}$  are ``left and right-handed'' 
sfermions).
As long as the low-energy theory does not contain a singlet which couples
to light fields, such terms do not destabilize the hierarchy.
However, their realization may require $\sqrt{F_{X}} \sim M \ll \mpl$,
\ie, a truly non-standard realization.
Other non-standard low-energy scenarios were also proposed.\cite{choi}

\section{Recent Models}
\label{sec:s3}

Equipped with the above ``catalog'' of operators and their manifestations,
we proceed to present two specific realizations in some more detail.
We choose two very different realization of the $n = 1$ tree-level
NMSSM operator,
the first in the context of gravity mediation while the second in the context
of gauge mediation. 

\subsection{Stabilized singlets and supergravity}
\label{sec:s31}

In Ref.~21 it was proposed that while a singlet field
with a VEV $\langle X \rangle \sim \sqrt{F_{X}} \sim m_{3/2}$
may induce $\mu$ and $m_{H_{UD}}^{2}$ in a supergravity scenario
with a (hidden-sector) spontaneous supersymmetry breaking scale
$\Lambda_{\rm SUSY} \sim 10^{11}$ GeV, its mass may be 
$m_{X} \sim {\cal{O}}(\Lambda_{\rm SUSY})$. This proposal
combines aspect of schemes based non-renormalizable supergravity operators
with some of the basic features of the NMSSM, and we will examine it 
in some detail.

As noted above, the NMSSM has two well-known problems.
At the renormalizable level, the NMSSM has a $Z_3$
symmetry. If that symmetry is preserved to all orders, then the VEV of $X$ will
break the symmetry at the weak scale and produce cosmologically dangerous
domain walls. If, on the other hand, the $Z_3$ symmetry is not preserved by
higher-order terms in the Lagrangian, then $X$ carries no conserved quantum
numbers. In this latter case, $X$ will generically
develop tadpoles, in the presence of spontaneously-broken
supersymmetry, whose quadratic divergences are cut
off by the Planck scale\cite{old,bp}. The resulting 
shift in the potential for $X$ causes it to slide to large values far
above the weak scale. If it were to couple to the MSSM Higgs fields,
they would receive unacceptably large masses, destabilizing the weak
scale. Therefore one concludes that not only do singlets fail to
provide a viable $\mu$-parameter, but they cannot even be allowed to
couple to light fields.

It is possible, however to solve this destabilization problem of the
NMSSM, while at the same time introducing a new
visible sector interaction whose scale will naturally fall at $\mint$.
Singlets can now couple to MSSM fields, and in
particular, can provide a dynamical $\mu$-term at the weak scale.
The model presented here
demonstrates a very general mechanism, and  it already contains all of the
ingredients necessary to be phenomenologically viable
(though these aspects will not be developed here).

Consider a superpotential
\beq
W=\lambda_H XH\bar H + \lambda_\Sigma X\Si\bar\Si
\eeq
where $H,\bar H$ carry charges $\pm1$ under a gauge symmetry $U(1)_H$, 
$\Si,\bar\Si$ are charged $\pm1$ under another gauge
symmetry $U(1)_\Si$, and $X$ is a gauge singlet. We require $\Si,\bar\Si$ to
be neutral under $U(1)_H$ and, for simplicity, assume that $H,\bar H$
are also neutral under $U(1)_\Si$, though they need not be. The gauge
$D$-term for $U(1)_\Si$ is then simply
\beq
D_\Si=g_\Si\left(|\Si|^2-|\bar\Si|^2\right),
\eeq
and similarly for $D_H$. To apply this toy model to the MSSM, we identify
$H,\bar H$ as the usual Higgs doublets, and extend 
$U(1)_H$ to the Standard Model
gauge group; $\Si$ and $\bar\Si$ are new fields charged under a new gauge
symmetry $U(1)_\Si$.

At the level of the superpotential, there exists the usual $Z_3$ which
forbids explicit mass terms from appearing in $W$\cite{nmssm}. 
This symmetry is broken
by $X\neq0$, which could lead to creation of 
electroweak scale domain walls (via the Kibble mechanism). 
The appearance of an $X^3$ term is
forbidden by an $R$-symmetry under which $R(W)=2$ and $R(X)=0$.
However, we will assume that the $Z_3$ symmetry of the superpotential is
only an accidental symmetry. This is a natural
expectation since global symmetries are generally not 
preserved by quantum gravity effects (unless they are remnants of 
broken gauge symmetries). 
In particular, we expect that gravity-induced global symmetry breaking will
appear as non-renormalizable, 
explicit symmetry-breaking terms in the K\"ahler potential. 
The $Z_3$ symmetry can thus be a symmetry of the effective
superpotential without being a symmetry of the entire action.
This is equivalent to the statement that the $X$ field is a true 
singlet, carrying no conserved quantum numbers. 

After supersymmetry breaking, non-zero tadpoles for $X$ 
will generically arise with 
light chiral fields circulating in the loops\cite{old,bp,jain,bpr,NP}. 
These tadpoles appear quantum mechanically
due to supergravity corrections from
Planck-suppressed operators. Because the exact source of the couplings which
generate the tadpoles is highly model-dependent, we do not know {\it a priori}
at what loop order non-zero contributions are generated. For example, it is
known that for a flat K\"ahler metric, non-zero tadpoles do not arise
until two-loops\cite{jain,bpr}; however, for a non-flat metric they may 
arise at one-loop.

The contribution to the effective (component field) potential 
coming from the tadpole  can be parameterized as
\be
V_{\rm linear} \sim
\beta\eps\mgrav\mpl F_X + \gamma\mgrav^2\mpl X + {\rm H.c.}
\label{eq:leff}
\ee
where $\eps$ is the maximum of a set of measures
$\eps_i$ of the supersymmetry breaking field VEVs (in Planckian units), 
and $\gamma,\beta$ are complex coefficients 
which include the loop suppression factors $(16\pi^2)^{-n}$  
(typically $n= 1$ or $n = 2$) as well as
counting factors $N$ which sum all unknown coefficients, 
and so whose magnitudes are roughly ${\cal O}(10^{-4}-1)$.
(We implicitly assume that $N$ is such that the calculation
remains perturbative, \ie, $N \lesssim 100$.)

Combining Eqs. (\ref{eq:leff}) and (\ref{V}) one can write
down the full scalar potential after supersymmetry breaking,
including supergravity-mediated soft masses as well as the tadpole
contributions. Begin by considering the contributions to the scalar
potential involving the $F_X$ auxiliary field:
\beq
V_{F_X}=(\beta\eps\mgrav\mpl F_X+{\rm H.c.}) -|F_X|^2
- \left(F_X\frac{\partial W}{\partial X}+{\rm H.c.}\right)
\eeq
where the first term is the contribution of the tadpole.
On integrating out all auxiliary fields, one finds that $F_X$
is shifted from its canonical form by the tadpole contribution:
\beq
F_X^\dagger=-\frac{\partial W}{\partial X} +\beta\eps\mgrav\mpl,
\label{FX}
\eeq
while all other $F$-terms (\eg\ $F_\Si$ and $F_{H}$)
are canonical. The $D$-terms associated with the
gauge fields also take their canonical forms.

The full scalar potential
after soft supersymmetry breaking can then be written:
\bea
V&=&\sum_i m_i^2|\varphi_i|^2
+|\lambda_\Si X|^2\left(|\Si|^2+|\bar\Si|^2\right) 
+|\lambda_H X|^2 \left(|H|^2+|\bar H|^2\right) \nonumber \\
& &{}+\mgrav^2\mpl(\gamma X+\gamma^\dagger X^\dagger)  
+\left|\lambda_\Si\Si\bar\Si+\lambda_H H\bar H-\beta\eps\mgrav\mpl\right|^2 \\
& &{}+\frac{g_\Si^2}{2}\left(|\Si|^2-|\bar\Si|^2\right)^2 
+\frac{g_H^2}{2}\left(|H|^2-|\bar H|^2\right)^2, \nonumber
\label{eq:V}
\eea
where the first term represents the
gravitationally-induced soft supersymmetry breaking masses, 
$m_i^2\sim\mgrav^2$, for the
fields $\varphi_i=\lbrace X,\Si,\bar\Si,H,\bar H\rbrace$, 
and the superpotential derivative in Eq.~(\ref{V}) was replaced
with the right hand side of Eq.~(\ref{FX}).
For simplicity,
we ignore hereafter holomorphic trilinear ($A$) and bilinear ($B$) SSB terms; 
they do not change our results
substantially. Note that 
the potential as written requires that $m_X^2\geq0$ in order to be bounded 
from below (this condition is modified in the presence of $B$-terms).
Indeed, one expects $m_X^2>0$ at tree level and it will only be
driven negative if 
its coupling to either of the two sets of Higgs fields is fairly large. 
Henceforth we will take all soft squared-masses to be equal to $\mgrav^2>0$.

To continue further, we take $\eps\simeq1$ which
is the generic choice; small deviations of $\eps$ away from 1 can be
absorbed into $\beta$.
Writing down the minimization conditions for the potential is straightforward,
but as the potential is quite complicated, it has many local 
minima besides the true
global one. However, there are two lowest-lying minima,
both along directions that are $D$-flat up to weak-scale corrections,
\ie, $\Si\simeq\bar\Si$ and $H\simeq\bar H$.

At a first minimum, denoted $V_1$, 
\bea
\Si=\bar\Si=H=\bar H&=&0, \nonumber \\
X&\simeq& -\gamma^\dagger\mpl, \nonumber \\
|F_X|&\simeq&|\beta\mgrav\mpl|, \nonumber \\
V_1\equiv V_{\rm min}&\simeq&(|\beta|^2-|\gamma|^2)\mgrav^2\mpl^2.\nonumber
\eea
This minimum represents the case usually considered in the literature for
singlets with non-zero 
tadpoles --- their VEVs are pulled up to the Planck scale, taking with them any
matter to which they couple. This is precisely the reason 
it was argued\cite{bp,bpr} that the
VEV of a true singlet cannot be responsible for the $\mu$-term in the MSSM.

At a second minimum, $V_2$, 
\bea
\Si\bar\Si&=&\frac{\beta\mgrav\mpl}{\lambda_\Si}, \nonumber \\
H=\bar H&=& 0,\nonumber \\
X&\simeq&-\frac{\gamma^\dagger}{2|\lambda_\Si\beta|}\mgrav, \nonumber \\
F_X&\sim&\mgrav^2, \nonumber \\
F_{\Si,\bar\Si}&\simeq&\lambda_\Si\mgrav^{3/2}\mpl^{1/2} ,\nonumber \\
V_2\equiv V_{\rm min}&\simeq&\frac{1}{|\lambda_\Si|}\left(\left|\beta\right|
-\left|\frac{\gamma^2}{2\beta}\right|\right)\mgrav^3\mpl. \nonumber
\eea
The $\Si$-fields receive
VEVs of $\sim\sqrt{\mgrav\mpl}$ to cancel off the $F_X$ contribution to the 
potential. These large $\Si$-VEVs then produce masses for the $X$-field
(through the $F_\Si$ terms) which stabilizes the $X$-VEV against the 
tadpole-induced linear potential. The resulting VEV of $X$ is then only
$\langle X\rangle\sim\mgrav\simeq\mw$!
Any gauge symmetry carried by
the $\Si$-fields will be broken at the scale of their VEVs. 
Up to the loop factors buried in $\beta$, this is the intermediate scale,
$\mint$. In fact, one
may interpret the physics at this minimum as the tadpoles communicating to the
$\Si$-fields the true scale of supersymmetry breaking, up to the loop factors.

There is also a third minimum, $V_3$, 
which is identical to $V_2$ except
that the would-be MSSM Higgs fields, $H$ and $\bar H$, 
play the role of $\Si$ and $\bar\Si$ and receive
VEVs $\sim\mint$, with $\lambda_H$ replacing $\lambda_\Si$ in all 
expressions. This is clearly {\it not}\/ the desired minimum but 
is instead another 
example of how the tadpole can destabilize the weak scale.
(Note that points at which $H,\bar H,\Si,\bar\Si$ all get VEVs
simultaneously are not even local minima of the potential.)

By observation, for $|\beta|^2>|\gamma|^2$ and  $|\lambda_H|<
|\lambda_\Si|$ $V_{2}$ corresponds to the global minimum and  
$X$ develops a VEV of the order of magnitude of 
the weak scale and provides a 
$\mu$-term of the correct size for light MSSM Higgs fields.
Note that simple inequality is all that 
is needed to ensure that a gauge hierarchy develops; no large 
hierarchy is needed between the two couplings themselves.
Unequal soft masses shift the condition slightly, but the same basic result
will always hold. Thus we conclude that the $X$-VEV can in fact provide the
$\mu$-term of the MSSM as long as $|\lambda_H|<|\lambda_\Si|$. 
The whole question of which gauge group is broken at the scale
$\mgrav$ and which at $\mint$ may rest entirely on the relative size of two
couplings ($\lambda_H$ and $\lambda_\Si$) whose ratio is generically 
${\cal O}(1)$!

It is well-known that models
with a dynamical $\mu$-term can contain a
Peccei-Quinn ($PQ$) symmetry which would be spontaneously broken and thus
create an unwanted axion at the weak scale. To examine this possibility,
promote the $PQ$-symmetry to the previously discussed $R$-symmetry under which $X$ is neutral and all other superfields are singly charged. 
However, $R(\beta\mgrav\mpl)=2$ explicitly breaks the symmetry and the
would-be axions are all given masses near the intermediate scale, rendering
them harmless. But there still remains a residual $PQ$ symmetry
in the MSSM Lagrangian. This too is explicitly broken, this time by
$F_X\sim\mw^2$, which generates a $m_{H_{UD}}^{2}$-term in the Higgs sector 
($\sim F^\dagger_XH\bar H$) and 
gives mass to the pseudoscalar Higgs/would-be axion.

In conclusion, a new mechanism for obtaining 
a weak-scale $\mu$-parameter 
by adding a total singlet in conjunction with a new gauge interaction and
its accompanying Higgs sector was presented. 
We have used the tadpoles endemic to models
with singlets to drive the breaking of the new symmetry at
the intermediate scale $\sqrt{\mw\mpl}$, to remove all vestiges of 
the singlet from the low-energy theory, and to render all would-be axions
harmless. The intermediate scale may be further associated 
with the scale of a symmetry breaking governing 
the mass scale right-handed neutrino\cite{MOH}, for example.
Such models may be described as a ``decoupled NMSSM'' in which
the NMSSM singlet is integrated out near the scale of supersymmetry breaking
but its traces remain in the low-energy theory. 
Further details on the model, its vacuum structure 
(at tree and one-loop levels), and details and
possible interpretation of the corresponding
intermediate-scale physics can be found in Ref.~21.

\subsection{Gauge non-singlets and gauge mediation}
\label{sec:s32}

Next, we turn to consider an explicit realization of the NMSSM
within the framework of gauge mediation.
Given the special form of the $\mu$-puzzle,
$m_{H_{UD}}^{2} = \mu \mmess \sim \mu \Lambda_{\rm SUSY} \gg \mw^{2}$,
there are two possible classes of solutions. 
Either $\mu$ and $m_{H_{UD}}^{2}$ both
arise at tree level so that their size is not determined by 
a (Yukawa) loop suppression factor raised to some power, 
or alternatively, they arise
at different loop orders and hence with different powers of
the suppression factor.  Realizations of these ideas, however,
are far from straightforward.

The most successful attempts to address this new hierarchy problem
fall along these lines and 
involve in one fashion or another the details of 
the high-energy (supergravity) theory, and in that sense
they are high-energy solutions. For example, one can invoke\cite{DGP} 
a radiative linear term generated by messenger-scale singlet interactions.
The linear term shifts a singlet field $X$ (which interacts with the
Higgs doublets) to a scale which is
suppressed by a loop factor in comparison to the messenger scale.
The shifts in the scalar and auxiliary
components of $X$, which induce $\mu$ and $m_{H_{UD}}^{2}$, respectively,
arise at different loop orders, evading the above described hierarchy
problem.  The superpotential (or equivalently -- the K\" ahler potential) 
couplings must be fixed by the  high-energy ($Q > \mmess$) theory. In
particular, a scale associated with a tree-level 
linear term must be fixed to be ${\cal{O}}(\mmess)$.
Alternatively, it was pointed out\cite{NP} 
that a radiative linear term
in a singlet field $X$ is typically generated by supergravity 
and is suppressed by only one inverse power of the Planck mass $M_{\rm P}$,
as demonstrated in Sec.~\ref{sec:s31} above.
Hence, it can still play an important role in the low-energy theory.
It shifts the singlet field $X \sim (\Lambda_{\rm SUSY}^{4}
/\kappa^{2}M_{\rm P})^{1/3}$
(assuming in this case $W(X) \sim (\kappa/3)X^{3}$ and 
$\epsilon \rightarrow 0$).
The singlet Yukawa interaction with the Higgs doublets then generates
the desired parameters at tree-level 
$\mu^{2} \sim m_{H_{UD}}^{2} \sim X^{2} \sim
\mw^{2}$ (assuming that supersymmetry is spontaneously broken 
at a scale $\mint \sim (4\pi/\alpha)\mmess \sim {\cal{O}}(10^{6\pm1})$ GeV). 
In this case no new scales are introduced by hand, but there is still
dependence on the high-energy theory. 
(A somewhat similar application of
supergravity to the problem was proposed in Ref.~40.)

Both proposals qualify as versions of a ``decoupled NMSSM''
in the sense described in Sec.~\ref{sec:s31} above, only that
the decoupling scale is now $\mmess$.
Here, however, we point out a distinctive possibility that the singlet 
field is not a gauge singlet but only a SM singlet $S$ which does
not decouple at the messenger scale (and hence will be denoted by $S$), 
\ie, a gauged NMSSM.
This possibility was discussed in Ref.~41, which we follow.
Specifically, let us assume
the extension $(S)SM$ $\rightarrow$ $(S)SM$ $\times$ $U(1)$$^{\prime}$, and  
that $S$ carries a charge $Q_{S} = - (Q_{H_{1}} + Q_{H_{2}})$ under
the  additional Abelian symmetry so that a Yukawa 
term $W\sim h_s S H_{1}  H_{2}$ is allowed. In turn, a scale 
$\Lambda^{\prime} \sim \langle S \rangle \lesssim \mmess$, which is
associated with the breaking of the $U(1)$$^{\prime}$, must be
introduced, or preferably, induced.
The $\mu$ and $m_{H_{UD}}^{2}$ parameters are induced
by the singlet interactions at tree-level and
the various $\mu$ problems of gauge mediation 
are solved in this case by the low-energy dynamics
associated with this new scale.

The scale $\Lambda^{\prime}$ could be  generated radiatively 
and is a function in this case of 
$\mmess$ and of ${\cal{O}}(1)$ Yukawa couplings. A coupling between $S$
and exotic quarks, e.g., $D$ and $D^{c}$ singlets with hypercharge
$\pm(1/3)$, generates negative
corrections to the SSB parameter $m_{S}^{2}$ 
so that $m_{S}^{2}(\Lambda^{\prime}) < 0$ and 
$S$ acquires a VEV. This is essentially a $U(1)$$^{\prime}$ version of the
well-known radiative symmetry breaking (RSB) mechanism that 
is responsible in the MSSM for the generation of 
the negative mass term in the SM Higgs potential
and the satisfaction of the conditions for EWSB discussed 
in Sec.~\ref{sec:pheno}. 
A similar idea\cite{prime} was 
mentioned above in the context of supergravity and high-energy
(gravity) mediation of supersymmetry breaking. 
In that case, like RSB in those models, 
the large evolution interval enables one to render 
$m_{S}^{2} < 0$ somewhere above the weak scale. 
In the supergravity case the superpotential interactions generate
$|\mu| \sim h_s \langle S \rangle$
while trilinear SSB terms
$V_{SSB} \sim \cdots + h_s A_s S H_{1}  H_{2} + {\rm H.c.} + \cdots$ generate
$m_{H_{UD}}^{2} = A_s h_s \langle S \rangle $. Since all parameters 
in the gravity-mediation framework
are of the same order of magnitude as the gravitino mass (which is
fixed in that case $m_{3/2}\sim \mw$),
then $h_s \langle S \rangle $ is expected 
to be of the same order of magnitude as well.
This leads to  a successful solution to the $\mu$-problem
in high-energy supergravity models. 
In contrast to the supergravity framework, 
in gauge mediation the evolution interval is short; in addition,
trilinear parameters are highly suppressed
$A \sim (\alpha/4\pi)^{2}\mmess\ln \mmess$ so that $m_{H_{UD}}^{2} \propto A\mu$
is also suppressed, even if $\mu$ is acceptable.
(Formally, $m_{H_{UD}}^{2}$ now arises at a too high loop order!)
While the small $A$ parameters remain a constraint,
the shorter evolution interval
is more than compensated (as for the case of RSB in these models) by the large
hierarchy within the SSB parameters
$m_{D}^{2}/m_{H}^{2}/m_{S}^{2} \sim
\alpha_{3}^{2}/\alpha_{2}^{2}/\alpha^{2}_{1^{\prime}}$
(where $m_{D}^2$, $m_{H}^2$ and $m_{S}^2$ 
are the soft mass-squares of the exotic
quark $D$, Higgs doublet $H$ and singlet
$S$ and $\alpha_{3,\,2,\,1,\,1^{\prime}}$ are the
$SU(3),\,SU(2),\,U(1)$, and $U(1)^{\prime}$ gauge couplings). In fact, the
messengers may not transform under $U(1)^{\prime}$, in which case 
$m_{S}^{2}(\mmess) = 0$. For $\alpha_{1^{\prime}} =
{\cal{O}}(\alpha_{Y})$, which we will assume,
the exact boundary condition for $m_{S}^{2}$ does not affect our discussion
and for simplicity we assume hereafter that the messengers are indeed
invariant under $U(1)^{\prime}$. 
(It can affect, however, the singlet slepton spectrum, 
which is otherwise given in gauge mediation only by hypercharge loops.)

A radiatively induced  $\langle S \rangle$ as a source of $\mu$
in the case of a gauge singlet $S$ was considered previously
in the context of gauge mediation.\cite{GMsinglet}
It was found that the singlet must couple to exotic quarks
with large Yukawa couplings, as naturally occurs in the context of
$U(1)$$^{\prime}$. In the gauge singlet case, however, the
superpotential must contain a $S^{3}$ term so that the potential contains
quartic terms $V \sim |\partial (S^{3} + SH_{1}H_{2}) / \partial S|^{2}$
which stabilize it.
Like the gravity-mediation versions of the NMSSM those models suffer from
the problem of a spontaneously broken global $Z_{3}$ symmetry 
(under which $S^{3}$ is invariant) which
results in unacceptable domain walls at a low-energy epoch.
In the gauged case $S$ is not a singlet and  
$S^{3}$ terms are not gauge invariant and are automatically
forbidden.  Instead, the potential is stabilized by $U(1)$$^{\prime}$ 
gauge $D$-terms $V \sim  \cdots + (g_{1^{\prime}}^{2}/2)(Q_{S}|S|^{2} +
Q_{H_{1}}|H_{1}|^{2}  + Q_{H_{2}}|H_{2}|^{2})^{2} + \cdots$
(which are not available for a gauge singlet $S$).
The $Z_{3}$ symmetry is now only a (harmless) subgroup of the gauged 
$U(1)^{\prime}$. 
While in the non-gauged case
the former source of the quartic terms also generates an additional
contribution to $m_{H_{UD}}^{2} \sim
S^{2}$, this is not possible in the gauged case 
(with only one singlet).

In either the gauged or non-gauged case, 
the potential also exhibits an approximate phase ($R$) symmetry, which
exists in models with only Yukawa superpotential terms and corresponds to a
rotation of all fields by the same phase. 
It is broken spontaneously by $\langle S \rangle$ and  explicitly by
tri-linear $A$-terms. The explicit breaking is, however,  suppressed by the
smallness of the $A$-parameters. Nevertheless, we will find below that 
in spite of the suppressed $A$-parameters 
it is possible to generate $m_{H_{UD}}^{2}$ 
and break the phase symmetry strongly enough
to avoid the light pseudo Goldstone boson which otherwise appears.
Specifically, as will be shown below, it is very likely that
in the $U(1)$$^{\prime}$ scenario $\mw \ll \langle S\rangle \lesssim
\mmess$ and hence, $m_{H_{UD}}^{2} \sim h_s A_s \langle S\rangle
\sim A_{s}\mu$ is a geometric mean of a small parameter and
a large VEV. It implies a somewhat  large 
value of $|\mu| \sim  {\cal{O}}(1\, {\rm TeV})$. However, this  
typically occurs in gauge mediation 
as a result of RSB constraints in the presence of a heavy gluino.
Alternatively, in models with two singlets a superpotential
term $SS^{\prime 2}$ could be gauge invariant, and
$\langle S^{\prime}\rangle \sim \langle S\rangle $
could generate an additional contribution to $m_{H_{UD}}^{2} \sim \langle 
S^{\prime}\rangle^{2}$, just as in the non-gauged case. 
(Note that in the non-gauged case 
the $U(1)^{\prime}$ rotations -- explicitly broken by the $S^{3}$
terms -- correspond to global transformations 
and there is one additional pseudo Goldstone boson.)
Here we confine ourselves to models with only one SM singlet $S$.

It is particularly interesting to note that in the models
with only one SM singlet there appear only two new phases which can be
rotated away, and hence there are no new physical phases
This is because there is only one common phase to all gaugino mass
and the radiatively-induced $A$ parameters, 
while the phase of $m_{H_{UD}}^{2}$ is given in this case 
by the phases of $\mu$ and $A$.
Hence, after $R$ and Peccei-Quinn rotations no physical phases
appear in the soft parameters.
This eliminates new
contributions to CP violating amplitudes such as the electron dipole
moment, which are flavor conserving and which generically appear at 
unacceptable levels even in gauge-mediation models.\cite{Moroi}

The stabilization due to the $D$-terms and the generation of the $A$
terms then open the door to new (low-energy) solutions to the
$\mu$-problem in gauge mediation. The mechanism is quite different from 
that of the non-gauged case since the quartic coupling is given, in
principle, by a fixed gauge coupling rather than by a free superpotential
coupling; $m_{H_{UD}}^{2}$ must depend on overcoming the suppression of the
tri-linear couplings $A$; and the scale $\langle S \rangle$ 
is a physical scale with observable consequences.
Hence, it corresponds to a distinctive and interesting option.
The $U(1)$$^{\prime}$ models predict, in addition to the extra matter
and the associated rich spectrum,
an extra gauge boson, $Z^{\prime}$. The corresponding phenomenology
is similar to that of any other model with $Z^{\prime}$, except that
$m_{Z^{\prime}}^{2} \sim -(Q_{S}/2)m_{S}^{2}$ is large,
given that $|m_{S}^{2}|$ is controlled by the large exotic quark
SSB parameters. Typically we find $m_{Z^{\prime}} \simeq {\cal{O}}(1
\, {\rm TeV})$ and with suppressed mixing with the ordinary $Z$-boson.
Thus it decouples safely from electroweak physics.
Another interesting aspect of supersymmetric $U(1)$$^{\prime}$ models
that repeats here is that the tree-level light Higgs $h_{1}$ 
mass exceeds its usual upper bound of $m_{Z}$.
This is due to contributions from the $U(1)$$^{\prime}$ 
$D$-terms to the quartic potential, which lift its otherwise flat 
direction. We find for its mass $m_{h_{1}} \simeq 120 - 150$ GeV 
at tree level and $m_{h_{1}} \simeq 150 - 180$ GeV at one loop.

A most interesting aspect of the $U(1)$$^{\prime}$ scenario is 
that the gauge-mediation scale is still the only
fundamental scale, and the $U(1)$$^{\prime}$ scale is determined
from it. It has been proposed recently\cite{CDM} that perhaps the same
$U(1)$$^{\prime}$ is also responsible for the actual mediation
of supersymmetry breaking from the ``hidden'' sector to the messenger fields 
(\ie, an ``active'' $U(1)$ whose primary role is to  
mediate supersymmetry breaking). 
This is an ambitious yet interesting proposal  that
significantly differs from our bottom-up  approach,  
which, in principle, is independent of the details of supersymmetry
breaking and its initial mediation to the messenger fields.
The $U(1)$ discussed here is a ``spectator'' (rather than ``active'')
$U(1)$ which does not participate
in the supersymmetry breaking or mediation mechanisms.
It witnesses supersymmetry breaking to the extent that the SM does
(or even less so if the messeneger fields are invariant under it).
By distinguishing the two extended interactions we avoid the need, 
e.g., to fine tune
Yukawa couplings, which is the situation in Ref.~44 due to the
multitude of tasks imposed there on a single  $U(1)$. 
The only (moderate) hierarchy in Yukawa couplings that is
assumed is between those that involve (exotic) quarks, which are
taken to saturate or be near their infra-red quasi-fixed points 
and be ${\cal{O}}(1)$, and those
which involve only the Higgs doublets and the singlet(s), which do
not reach any (quasi-)fixed points and hence are taken to be smaller.
Such differences naturally stem from QCD renormalization, which enables
the existence of quasi-fixed points for the (exotic) quark couplings.

We now turn to describe a specific model in detail.
The superpotential reads
\begin{equation}
\label{w}
\begin{array}{cc}
W = & 
- h_{U}H_{U}QU + h_{D}H_{D}QD+ h_{E}H_{D}LE \\
 & -\lambda_{S}S H_{1}  H_{2} + \lambda_{D} S D_{i} D^{c}_{i},
\end{array}
\end{equation}
where the MSSM has been extended to include the singlet $S$ and exotic 
quark vector-like pairs $D$ and $D^{c}$ 
which are singlets under $SU(2)$$_{L}$ and carry hypercharge $\pm 1/3$.
In Eq.~(\ref{w}) we include the usual Yukawa terms involving the third
generation fields, an effective $\mu$-term $\lambda_{S}S H_{1}  H_{2}$,
and a Yukawa coupling between the singlet and the exotic quark superfields. 
Given our assumptions,
the free parameters in the analysis are $\mmess$, the number $n_{D}$ of $D$,
$D^{c}$ pairs that couple to $S$, and the corresponding  Yukawa couplings
$\lambda_{D}$. 
($\lambda_{S}$ is fixed by the minimization condition (\ref{MZ}).) 
The product of $n_{D}$ and $\lambda_{D}$ is constrained by electroweak
breaking and also by requiring a sufficiently heavy $Z^{\prime}$.
Below we fix, as an example, 
$n_{D} = 3$ and $\lambda_{D} = 0.7$. 
Following Eq.~(\ref{V}),
the most general renormalizable potential involving the Higgs
singlet field $S$ is
\begin{eqnarray}
V& =& |\lambda_{S} H_{1} H_{2}|^{2}+
|\lambda_{S}S|^{2} (|H_{1}|^{2}+|H_{2}|^{2}) \nonumber \\
& +& \frac{G^{2}}{8}(|H_{1}|^{2}-|H_{2}|^{2})^{2} +
\frac{g^{2}}{2}|H_{1}^{\dagger}H_{2}|^{2} +
\frac{g_{1^{\prime}}^2}{2}(Q_{1}|H_{1}|^{2}
  +Q_{2}|H_{2}|^{2}+Q_{S}|S|^{2})^{2} 
\nonumber \\
&+& m_{H_{D}}^{2}|H_{D}|^{2} + m_{H_{U}}^{2}|H_{U}|^{2} + m_{S}^{2}|S|^{2}
 + ( A_{s} \lambda_{S} S H_{1}  H_{2} + {\rm H.c.}), 
\label{vprime} 
\end{eqnarray} 
where $G^{2}=g_{Y}^{2}+g^{2}$ and the the SSB terms are listed in last line.

The experimental constraint on the mass of the $Z^{\prime}$ 
can be satisfied if 
the $U(1)$$^{\prime}$ is broken at the TeV scale, 
which requires $\langle H_{D} \rangle, \langle 
H_{U} \rangle \ll \langle S \rangle,\,\langle S^{\prime}
\rangle$. This separation is indeed realized in our example and
the determination of the $S$ VEV can therefore be separated 
to a very good approximation from that of 
the Higgs VEV's.  The scalar potential for $S$ then reads 
\begin{equation} 
\label{vs}
V= m_{S}^{2}|S|^{2} + \frac{g_{1^{\prime}}^2}{2} (Q_{S}|S|^2)^{2}. 
\end{equation} 
It acquires a VEV  $\langle S \rangle = s/\sqrt{2}$ where
\begin{equation} 
\label{s}
s^{2}= - \frac{2 m_{S}^{2}}{g_{1^{\prime}}^{2}Q_{S}^{2}}, 
\end{equation} 
if the evolution of $m_{S}^{2}$ can be neglected near the minimum. Hence, a
large value for $s$ occurs for
$m_{S}^{2}$ large and negative. This is achieved by the
order unity Yukawa couplings between $S$ and exotic quark pairs $D$
and $D^{c}$ (with scalar mass-squares $m_{D,\,D^{c}}^{2}(\mmess) \gg
m_{S}^{2}({\mmess}) \simeq 0$), 
which rapidly diminish $m_{S}^{2}(Q < \mmess)$ via the usual
renormalization group evolution.  
The mass of the $Z^{\prime}$ boson, which is 
independent of $g_{1^{\prime}}$, is
\begin{equation}
m_{Z^{\prime}} \sim g_{1^{\prime}} Q_{S} s \sim \sqrt{2|m_{S}^{2}|},
\end{equation} 
with the $Z-Z^{\prime}$ mixing angle $\alpha_{Z-Z^{\prime}} = {\cal O}
(m_{Z}^{2}/m_{Z^{\prime}}^{2})$.   The $Z^{\prime}$ mass 
and the $U(1)$$^{\prime}$ scale are  determined by
the only scale in the problem, $\mmess$ (which is encoded in $m_{S}^{2}$). 
The VEV of $S$ generates an effective $\mu$-parameter
$\mu=\lambda_{S}s/\sqrt{2}$. The $A$-term associated with $S H_{D}  H_{U}$,
which is non-zero at the electroweak scale due to gluino loop corrections, 
generates an effective $m_{H_{UD}}^{2}$ for the two Higgs doublets 
$m_{H_{UD}}^{2}= A_{s}\mu$. 
(In addition, the $U(1)$$^{\prime}$ $D$-term 
generates corrections to the Higgs
scalar masses $\delta m_{H_{D,\,U}}^{2} = 
({g_{1^{\prime}}^2}/{2})Q_{1,\,2} Q_{S} s^2$.)

As an example of actual derivation of RSB and the spectrum, 
let us list the model spectrum at the infra-red $\sim \mw$
in the case of the $E_{6}$ $U(1)_{\eta}$ assignments 
$Q_{1} = 1$, $Q_{2} = 4$ and $Q_{S}=-Q_{1}-Q_{2}$,
$\mmess = 10^5$ GeV, three pairs of exotic quark singlets, 
and $\lambda_{D}(\mmess) = 0.7$. (For full listing and other examples, see
Ref.~41.)
The VEV of the singlet is $s=3720~{\rm GeV}$, 
the VEV's of the Higgs doublets are
$\langle H_{D}^{0}\rangle =14~{\rm GeV}$ 
and $\langle H_{U}^{0}\rangle = 245~{\rm GeV}$, 
resulting in a solution with $\mu = 1050~{\rm GeV}$ 
(or equivalently, $\lambda_{S}(\mmess) = 0.47$)
and $\tan{\beta}=18$. The effective $m_{H_{UD}}^2$ is $\sim
(235\,{\rm GeV})^2$. The $Z^{\prime}$ mass is $m_{Z^{\prime}}=1110$ GeV  and  
the $Z-Z^{\prime}$ mixing angle is $\alpha_{Z-Z^{\prime}} =
0.004$. The  (tree-level) spectrum of the
CP even physical Higgs is $m_{h_{1}}=124~{\rm GeV}$, 
$m_{h_{2}}=995~{\rm GeV}$,
$m_{h_{3}}=1090~{\rm GeV}$, while $m_{h_{1}}= 154~{\rm GeV}$ at one loop
(with negligible corrections to $m_{h_{2,\,3}}$).
The CP odd Higgs scalar 
and the charged Higgs masses are $m_{A} \simeq m_{H^{\pm}}=993~{\rm GeV}$. The heaviest
CP even Higgs scalar $h_{3}$ is mainly composed of the singlet $S$, associated with the
breaking of the $U(1)$$^{\prime}$. The second heaviest CP even Higgs, 
the CP odd
Higgs and the charged Higgs fields form the $SU(2)$ doublet that is not
associated with the ${SU(2)}\times {U(1)}_Y$ breaking.   
The masses of the two charginos are 
$m_{\tilde{\chi}_{1}^{\pm}}=266~{\rm GeV}$ and
$m_{\tilde{\chi}_{2}^{\pm}}=1060~{\rm GeV}$. 
The lightest (heaviest) chargino is
predominantly a gaugino (Higgsino). The spectrum of the neutralinos is
$m_{\tilde{\chi}_{1}^{0}}=142~{\rm GeV}$, 
$m_{\tilde{\chi}_{2}^{0}}=266~{\rm GeV}$, 
$m_{\tilde{\chi}_{3}^{0}}=1060~{\rm GeV}$, 
$m_{\tilde{\chi}_{4}^{0}}=1060~{\rm GeV}$, 
$m_{\tilde{\chi}_{5}^{0}}=1120~{\rm GeV}$, 
$m_{\tilde{\chi}_{6}^{0}}=1120~{\rm GeV}$. 
In the limit of neglecting the Higgs VEVs, the two lightest
neutralinos are just $\tilde{B}$ and $\tilde{W_3}$, \ie, the 
Bino and the Wino. $\tilde{\chi}_{3,\,4}$ are linear 
combinations of Higgsinos with 
nearly degenerate masses $\sim \mu=h_s s/\sqrt{2}$; 
and  $\tilde{\chi}_{5,\,6}$
are linear combinations of the other gaugino 
$\tilde{B}^{\prime}$ and the singletino $\tilde{S}$ 
with degenerate masses $\sim m_{Z^{\prime}}$.
Squark and gluino masses are in the $1200 - 1400$ GeV
range. The next to lightest sparticle (NLSP) is the lightest neutralino, 
which  is predominantly the bino, \ie, the gaugino of the $U(1)_Y$.

We note in passing 
the usual near equality between $|\mu|$ and the 
gluino mass that often appears 
in various variants of the MSSM. (See Sec.~\ref{sec:pheno} for a discussion.)
The gluino is heavy, which is a  generic prediction of the gauge-mediation
framework, and hence it naturally leads to a relatively
large value of $\mu$. In particular, there is no new tuning 
due to  the $U(1)$$^{\prime}$ dynamics. 
The fine-tuning question can be phrased in this case in terms
of the particular value of $\lambda_{S}$.
It is worth stressing that the Higgs
mixing parameter in the scalar potential $m_{H_{UD}}^{2} = \mu A_{s}$ is a
geometrical mean of the superpotential Higgs mixing parameter $\mu$
and a radiatively generated (small) trilinear coupling $A_{s}$. Since
$\mu$ is proportional to a large VEV
(and the  heavy gluino implies further that $|\mu|$ is not suppressed by a
small coupling) the geometrical mean $m_{H_{UD}}
\sim {\rm a}\,\,{\rm few}\,\times\, 100$ GeV is sufficiently large.

In conclusion,
the Higgs mass parameters in the gauge-mediation framework are best
understood as dynamical degrees of freedom corresponding to a (SM) singlet.
Here, it was suggested that such a singlet is not a gauge singlet but
transforms under a $U(1)$$^{\prime}$.
The  $U(1)$$^{\prime}$ scale may be
generated naturally and radiatively one or two orders of
magnitude below the messenger scale. Upon integrating out the  
$U(1)$$^{\prime}$ sector, the supersymmetry conserving ($\mu$) and
breaking ($m_{H_{UD}}^{2}$) dimensionful Higgs mixing parameters
are generated, resolving the $\mu$ problem in the otherwise
attractive class of gauge-mediation models. 
The $U(1)$$^{\prime}$ dynamics also adds to the already strong
predictive power of the gauge-mediation framework, as the scalar,
fermion and vector electroweak sectors are extended and new exotic
matter is predicted at a few TeV scale. 
Further discussion and examples can be found in Ref.~41.

\section{Generalization to Lepton Number Violation}
\label{sec:s4}

Before concluding, let us note that
the MSSM contains three more gauge-invariant
holomorphic bilinear terms, $L_{i}H_{U}$ $i=1,\,2,\,3$,
which could appear in the superpotential or the scalar potential,
corresponding to lepton number violating (LNV) $\mu$ and $m_{H_{UD}}^{2}$
(or $B$) terms. In the MSSM one imposes the conservation
of $R$-parity, $R_{P} = (-)^{3B + L + 2S}$, which encompasses both
lepton $L$ and baryon $B$ number conservation 
(and $S = 0,\,1/2,\,1$ is the particle spin).
This is sufficient but not necessary to ensure proton stability.
LNV terms could be admitted as long as baryon number is 
(sufficiently) conserved!
It was proposed\cite{HS} that $W_{\rm LNV} \sim \mu_{L}LH_{U}$ 
could therefore be present and provide an electroweak scale source
for neutrino masses and mixing.

This brings us back to the operators (\ref{mu}).
By interplay of symmetries (which distinguish Higgs and lepton fields
in the ultra-violet theory) and $F$- and $D$-type  operators in Eq.~(\ref{mu}),
one can realize simultaneously both the usual and LNV $\mu$-terms 
once integrating out the background fields $X$.\cite{NP1}
They could be comparable in size or maintain  a certain hierarchy.
This offers, on the one hand, a minimal LNV extension of the MSSM,
while on the other hand, already contains an explanation to its minimality:
If LNV arises from non-renormalizable operators suppressed by some scale $M$,
then  on dimensional grounds LNV Yukawa couplings $h_{\rm LNV} \sim
\mu_{LNV}/M \rightarrow 0$, leaving the bilinear term as the primary source
of LNV. (Another possible explanation will be offered below.)

In theories in which the SSB parameters do not distinguish 
$H_{D}$ from the (s)lepton fields (Higgs-lepton universality), it is 
straightforward to show that in the infra-red, after minimizing the potential
and redefining the  (MSSM) Higgs $H_{D}$ along the VEV 
(in a four dimensional field space spanned by $H_{D}$ 
and the three lepton doublets),  $|\mu|$ is again
concentrated in the usual Higgs mixing term and there is no LNV
in the effective tree-level theory. However, quantum corrections 
proportional to the usual Yukawa couplings spoil the Higgs-lepton
universality, and therefore a small LNV remains in the renormalized theory
and it controls the size of the effective LNV Yukawa couplings
(which are given by the usual Yukawa couplings times a rotation angle)
as well as  the Higgsino-neutrino mixing, and hence the neutrino spectrum.
This is the dynamical alignment mechanism\cite{NP1,Hemp} which asserts
that given Higgs-lepton universality the size of the neutrino mass
arising at tree-level from Higgsino-neutrino mixing is suppressed
by the dynamical alignment between the $\mu$ and VEV vecors in the 
relevant four-dimensional field space. 
The alignment itself arises trivially upon minimization
as the $\mu$ vector defines the only direction in field space;
hence, dynamical alignment. 
The extent of the alignment is determined by the relative size 
of the radiative corrections to the relevant SSB parameters,
which spoil the alignment.
These corrections can be calculated in a given model and typically
one finds neutrino masses of the order of MeV or smaller.
In fact, one can formulate these arguments in terms of a chiral
$SU(4)$ symmetry controlling the Higgsino-neutrino mixing which is
broken at tree level by the fundamental $\mu$ and VEV four-vectors
to either $SU(3)$, if the alignment holds 
(massive Higgsino but massless neutrino), 
or $SU(2)$ otherwise (massive Higgsino and one massive neutrino).

Having discussed a gauged NMSSM model in Sec.~\ref{sec:s32}, we note that
it is straightforward  to extend the discussion there
to include bilinear LNV through couplings
$h_{\not{L}}SLH_{U} \rightarrow \mu_{LNV}LH_{U}$ if $L$ and $H_{D}$
carry the same $U(1)$$^{\prime}$ charge (this is the case for a
$U(1)$$_{\eta}$ of $E_6$ embedding),
or more generally in a multi-singlet model. 
The  case $Q_{H_{D}} \neq Q_{L}$ is in fact more
attractive since it would forbid lepton number violating Yukawa operators 
in the high-energy theory.
Since gauge mediation guarantees Higgs-slepton
mass universality, and Higgs-slepton bilinear mixing in the scalar
potential arises only from radiative  $A$-parameters, then all
conditions for the dynamical alignment suppression of neutrino masses
are automatically and naturally 
satisfied. Hence, such an extension provides another realization of
the bilinear LNV framework.

LNV Yukawa couplings appear in the infra-red as a result of the
redefinition of the Higgs field along the VEV. As mentioned above,
they are proportional to the usual Yukawa matrices, which leads to a clear
prediction that the LNV Yukawa couplings in 
$W \sim \lambda_{i33}L_{i}Q_{3}D_{3}$ are proportional to $b$-quark
mass, and hence are the dominant LNV Yukawa couplings.
(This can have a significant effect on the corresponding
collider phenomenology,\cite{333} for example.) 
Since the usual Yukawa couplings do not respect the $SU(4)$ symmetry
(and hence lead to the radiative corrections to the alignment mentioned below),
then the LNV Yukawa couplings, which are proportional to the usual ones, also 
break  (maximally) the $SU(4)$ 
symmetry and lead to finite quantum corrections
to the neutrino masses so that all neutrinos
are massive at the quantum level.
This results in a variety of models and neutrino mass patterns.
It is a miraculous complementarity
between the SM where the neutrino must remain massless
and the MSSM where only the neutrinos can have an explicit mass term.
For that matter, Higgs and lepton fields are not distinguished
by the symmetries of the model, unless one imposes such a distinction.
It is therefore also crucial that one can naturally
generate a large hierarchy between the Higgsino and neutrino masses.
Phenomenology and viability of these models 
was studied extensively\cite{BRPV} over the last few years.

\section{Summary}
\label{sec:s5}

In summary, we have demonstrated the phenomenological correlation
between the size of the supersymmetry conserving $\mu$ parameter
and the SSB parameters and established rough lower and upper bounds 
for $|\mu|$. The $\mu$ puzzle was then defined as the question of the origin
of this correlation, which also suggests that the Higgs fields may be 
distinguished from all other matter in the way that they communicate
with heavy and supersymmetry breaking fields (labeled as the background fields
$X$ in this lecture).  
In order to address this puzzle in some generality
we pursued an operator analysis which was then applied
to different frameworks for the SSB parameters, 
comparing all possibilities on equal footing.
While in models with $m_{3/2} \sim \mw$ the correlation could
naturally arise from various sources, in models where
the SSB parameters and the weak scale are determined
by (gauge or gravity) quantum corrections
the puzzle transformed into a new hierarchy problem
between the Higgs mixing in the superpotential and the scalar potential.
Two specific frameworks for the solution, 
both based on unconventional variants of the NMSSM,
were then presented in some detail, the first assuming gravity mediation
and the second assuming gauge mediation of the SSB parameters.
Lastly, we also noted that $(i)$ $\mu$ may be a SSB parameter
in certain situations and $(ii)$ that it is straightforward to admit
Higgs-lepton mixing which generalizes the usual $\mu$-term and leads
to neutrino masses and mixing. The $\mu$-parameter and its mysterious origin
have fueled many  works in recent years. Here we attempted to catalog
some of those ideas and proposals, and elaborated on two more recent ones.
To conclude, the discovery of supersymmetry would lead to the measurement
of $\mu$, and the information it encodes could be unfolded.
The measurement and its interpretation
will then open a new window to the ultra-violet.

I would like to thank Professor J.~E.~Kim for his kind invitation
and the Korea Institute for Advanced Study and the Center for Theoretical
Physics at Seoul National University for their hospitality. Work supported
by the US Department of Energy under cooperative research
agreement No.~DE--FC02--94ER40818.
The support of Rutgers University 
during earlier parts of this research is acknowledged.

\end{document}